\documentclass{kapproc} % Computer Modern font calls
\usepackage{epsfig}

\kluwerbib
\begin{document}

\articletitle{Gamma Ray Pulsars}

\articlesubtitle{Multiwavelength Observations}

\author{David J. Thompson}
\affil{Laboratory for High Energy Astrophysics\\NASA Goddard Space Flight Center\\Greenbelt, Maryland 20771 USA}
\email{djt@egret.gsfc.nasa.gov}

\begin{keywords}
Pulsars, gamma rays, observations, multiwavelength
\end{keywords}

\begin{abstract}
High-energy gamma rays are a valuable tool for studying particle acceleration and radiation in the magnetospheres of energetic pulsars.  The seven or more pulsars seen by instruments on the Compton Gamma Ray Observatory (CGRO) show that: the light curves usually have double-peak structures (suggesting a broad cone of emission); gamma rays are frequently the dominant component of the radiated power; and all the spectra show evidence of a high-energy turnover.  For all the known gamma-ray pulsars, multiwavelength observations and theoretical models based on such observations offer the prospect of gaining a broad understanding of these rotating neutron stars.  The Gamma-ray Large Area Space Telescope (GLAST), now in planning for a launch in 2007, will provide a major advance in sensitivity, energy range, and sky coverage. 
\end{abstract}

\section*{Introduction}
%\[
%\widehat{a} + \widehat{ab} + \widehat{abc} + \widehat{abcd}
%\]
%
%%\show\frak
% 
%\[
%%      {\bf x}^{\bf x} \triangleq z 
%      {\bf x}^{\bf x}\triangleq{z} \tensor{T} \frak{E^E}=\frak{mc}^2
%%      {\bf x}^{\bf x}\triangleq {z} \tensor{T} \frak{E}=\frak{mc}^2
%\]
% 
%\[
%{\Bbb {NQRZ}} \qquad \because \eth\ggg\bigstar \therefore\blacktriangleright\rightsquigarrow \blacksquare
%\]
% 

Pulsars are particularly interesting astrophysical objects because we know so 
much about them. Most of that information is derived from their timing 
properties.     Measurement of the period P and period derivative $\dot{P}$ first identified 
pulsars as compact objects.  Under fairly general assumptions that pulsars are rapidly-rotating neutron stars with a strong dipole magnetic field, a whole range of physical 
parameters can be estimated from these timing parameters.	Examples include the 
timing age, the spin-down energy loss, the surface magnetic field, and the open field 
line voltage (Thompson, 2000).

\section*{Gamma-Ray Pulsar Multiwavelength Light Curves}
							
The telescopes on the Compton Gamma Ray Observatory identified seven or more gamma-ray pulsars, some with very high confidence and others with less certainty.    Figure 1 shows the light curves from the seven highest-confidence gamma-ray pulsars in five energy bands: radio, optical, soft X-ray ($<$1 keV), hard X-ray/soft gamma ray ($\sim$10 keV - 1 MeV), and hard gamma ray (above 100 MeV).  Based on the detection of pulsations, all seven of these are positive detections in the gamma-ray band. The weakest (PSR B1951+32) has a statistical probability of occurring by chance of $\sim$10$^{-9}$.

\begin{figure}[b!] % fig 1
\centerline{\epsfig{file=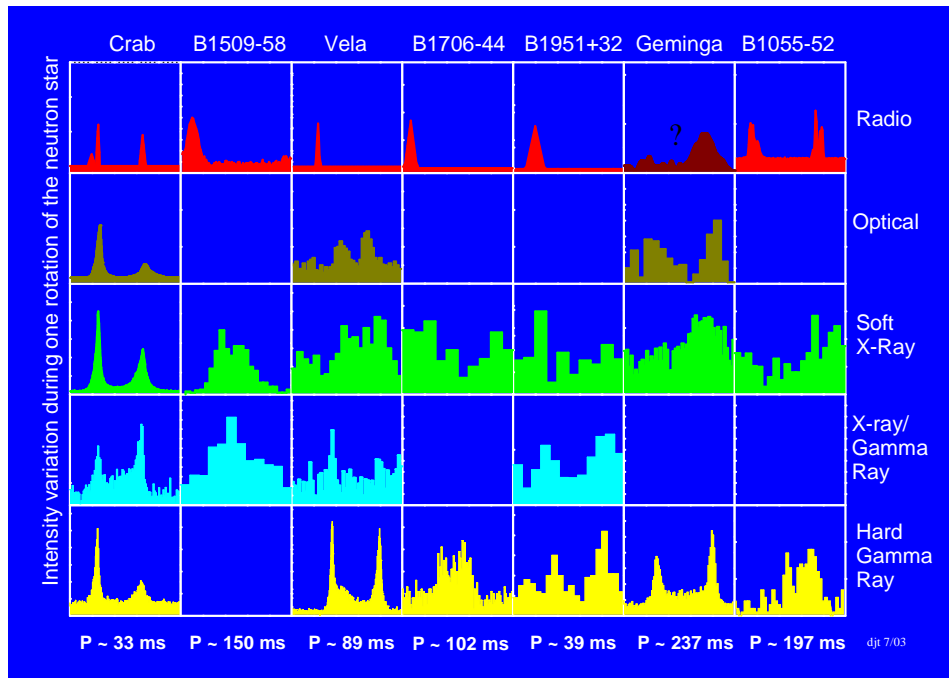,width=4.8in}}
\vspace{10pt}
\caption{Light curves of seven gamma-ray pulsars in five energy bands,  from left to right in order of characteristic age. Each panel shows one full rotation of the neutron star. The X-ray and gamma-ray references are those from Fig. 1 of Thompson (2001), with the addition of the X-ray data for PSR B1706$-$44 from Gotthelf, Halpern, and Dodson (2002).  Radio references (left to right): Manchester, 1971; Ulmer et al., 1993; Kanbach et al., 1994; Johnston et al., 1992; Kulkarni et al., 1988, Kuz'min and Losovskii,  1997; Fierro et al., 1993. Optical references: Crab, Groth, 1975; Vela, Manchester et al., 1980; Geminga, Shearer et al., 1998.}
\label{fig1}
\end{figure}

Some important features of these pulsar light curves are:
\begin{itemize}
\item 
 They are not the same at all wavelengths.  Some combination of the geometry and the emission mechanism is energy-dependent.  In soft X-rays, for example, the emission in same cases appears to be thermal, probably from the surface of the neutron star;  thermal emission is not the origin of radio or gamma radiation.

\item Not all seven are seen at the highest energies.  PSR B1509$-$58 (which has the strongest magnetic field among the gamma-ray pulsars) is seen up to 10 MeV by COMPTEL (Kuiper et al. 1999), but not above 100 MeV by EGRET.  

\item The six seen by EGRET all have a common feature - they show a double peak in their light curves.  Because these high-energy gamma rays are associated with energetic particles, it seems likely that the particle acceleration and interactions are taking place along a large hollow cone or other surface.  Models in which emission comes from both magnetic poles of the neutron star appear improbable in light of the prevalence of double pulses. 

\end{itemize}

\epsfclipon
\begin{figure}[b!] % fig 2
\centerline{\epsfig{file=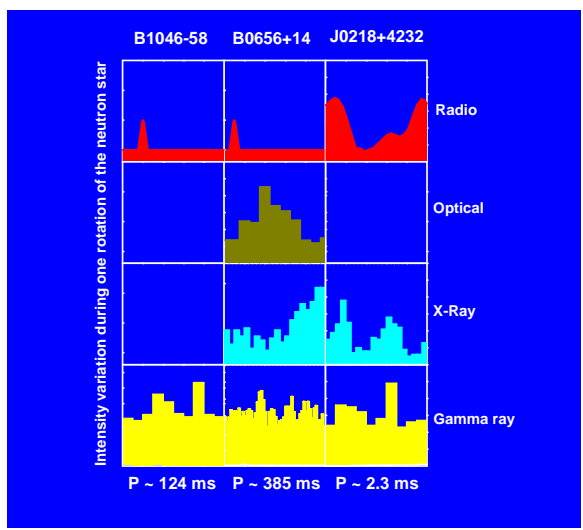,width=3in}}
\vspace{10pt}
\caption{Light curves of three candidate gamma-ray pulsars. Each panel shows one full rotation of 
the neutron star.  References: PSR B1046-58: (Kaspi et al. 2000); PSR B0656+14: (radio: Kuz'min et al. 1998; optical: Shearer et al. 1997; X-ray: Finley 1994, gamma-ray: Ramanamurthy et al. 1996); PSR J0218+4232 (Kuiper et al. 2000, 2002).}
\label{fig2}
\end{figure}

In addition to the six high-confidence pulsar detections above 100 MeV, at least three additional radio pulsars may have been seen by EGRET.  Figure 2 shows their light curves in radio, optical, X-ray, and gamma rays.  The gamma-ray light curves are shown without the zero suppression used in some of the original references.  These three all have statistical probabilities in the 10$^{-4}$ range, or about 5 orders of magnitude less convincing than the weakest of the seven on the previous figure.  These are good candidates, but they are not strong enough to be used as discriminators between models.  

Some features of these pulsars are:
\begin{itemize}
\item 
 PSR B1046$-$58, which may be the counterpart of 3EG J1048$-$5840, has properties similar to Vela and PSR B1706$-$44 (Kaspi et al. 2000).  It also has an associated X-ray source, but this source is not pulsed. 
 
\item PSR B0656+14 has timing properties similar to Geminga.  It is a well-known X-ray pulsar, and optical pulsations have been seen.  It is not detected in the EGRET catalog as a source by spatial analysis (Ramanamurthy et al. 1996). 

\item PSR J0218+4232 is the only millisecond pulsar with evidence of gamma-ray emission (Kuiper et al. 2000, 2002).  The observed gamma-ray pulse resembles that seen in X-rays. Spatial analysis of this gamma-ray source is complicated by the presence of 3C66A, a BL Lac object about 1 degree away from the pulsar on the sky (Hartman et al. 1999). 

\end{itemize}

{\bf A note on gamma-ray pulsar variability:}  although the pulsations themselves are a periodic variability with very high contrast between peaks and valleys, no strong evidence has been found for longer-term variability of the gamma-ray emission from pulsars, in contrast to many other gamma-ray sources (e.g. McLaughlin et al. 1996, Torres et al. 2001, Nolan et al., 2003).

\begin{figure}[b!] % fig 3
\centerline{\epsfig{file=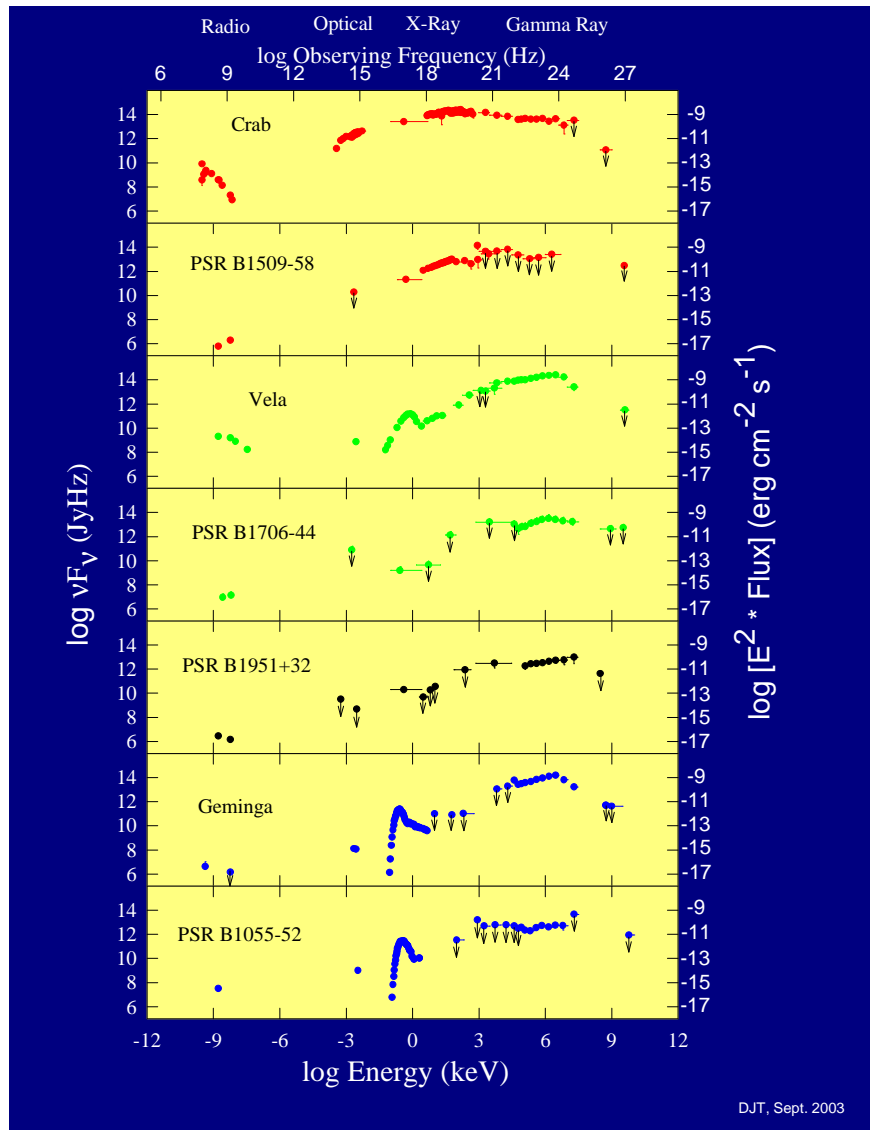,width=4.5in}}
\vspace{10pt}
\caption{Multiwavelength spectra of seven gamma-ray pulsars.  Updated from Thompson et al. (1999) with data from Fierro et al. (1998) for the Crab (gamma rays) and Jackson et al. 2002 for Geminga (hard X-rays). }
\label{fig3}
\end{figure}

\begin{figure}[b!] % fig 4
\centerline{\epsfig{file=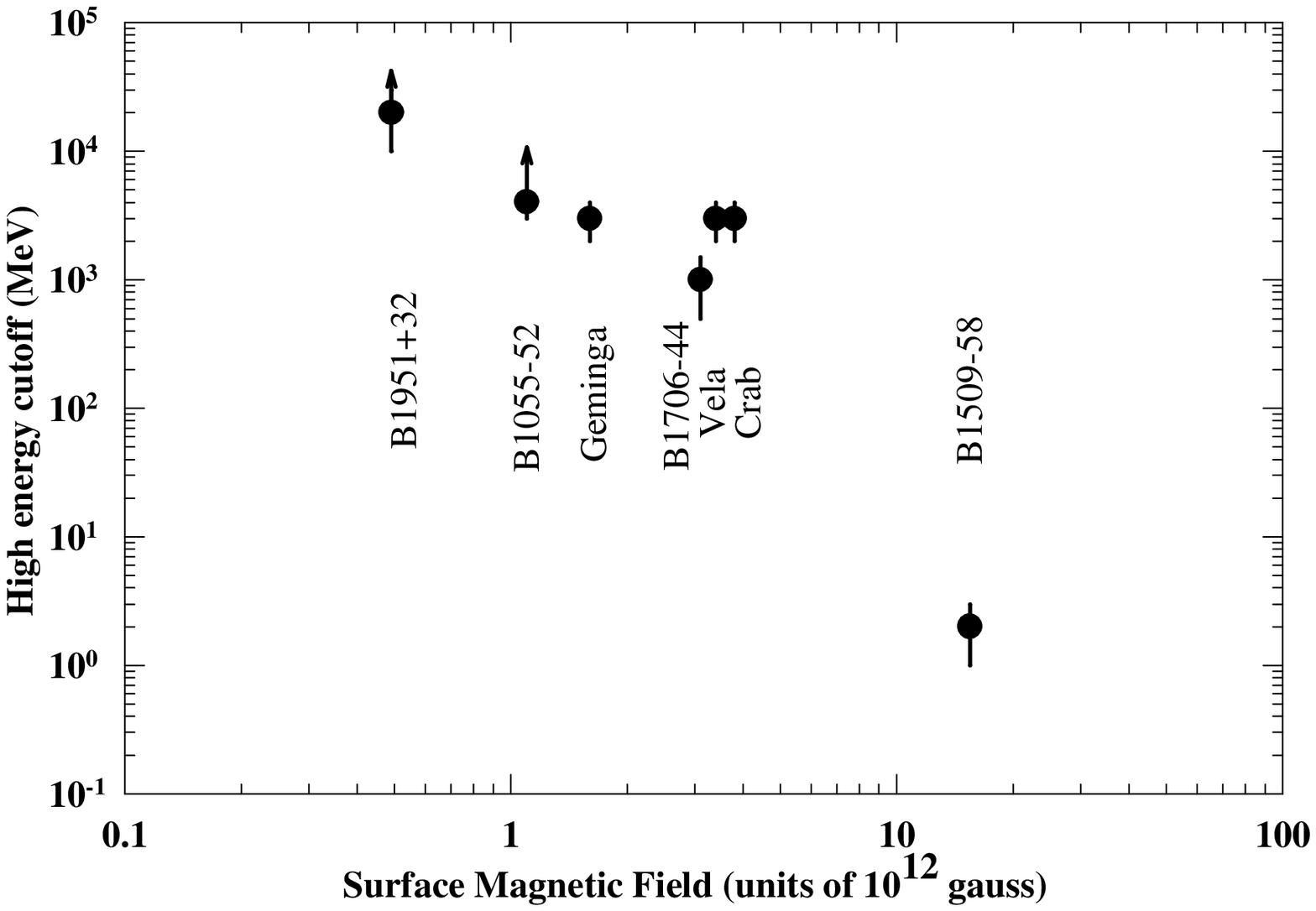,width=4in}}
\vspace{10pt}
\caption{High-energy cutoff energy for gamma-ray pulsars as a function of surface magnetic field.  For PSRs B1951+32 and  B1055$-$52, only a lower limit to the cutoff energy can be determined from the EGRET data. }
\label{fig4}
\end{figure}

\section*{Gamma-Ray Pulsar Multiwavelength Energy Spectra}

While the pulsations identify sources as rotating neutron stars, the observed energy spectra reflect the physical mechanisms that accelerate charged particles and help identify interaction processes that produce the pulsed radiation.  Broadband spectra for the seven highest-confidence gamma-ray pulsars are shown in Fig. 3.  The presentation in $\nu$F$\nu$ format (or E$^2$ times the photon number spectrum) indicates the observed power per frequency interval across the spectrum.  In all cases, the maximum power output is the in the gamma-ray band. Other noteworthy features on this figure are:

\begin{itemize}
\item 
 The distinction between the radio emission (which originates from a coherent process) and the high-energy emission (probably from individual charged particles in an incoherent process) is visible for some of these pulsars, particularly Crab and Vela. 
 
\item Vela, Geminga, and B1055$-$52 all show evidence of a thermal component in X-rays, thought to be from the hot neutron star surface. 

\item The gamma-ray spectra of known pulsars are typically flat, with most having photon power-law indices of about 2 or less.  Energy breaks are seen in the 1-4 GeV band for several of these pulsars.  These changes in spectral index appear to be related to the calculated surface magnetic field of the pulsar, as shown in Fig. 4.  The lowest-field pulsars have no visible break in the EGRET energy range; the existence of a spectral change is deduced from the absence of TeV detections of pulsed emission.  The highest-field pulsar among these, B1509$-$58, is seen only up to the COMPTEL energy band.

\item No pulsed emission is seen above 30 GeV, the upper limit of the Compton Observatory/EGRET observations.  The nature of the high-energy cutoff is an important feature of pulsar models. 

\item Although Fig. 3 shows a single spectrum for each pulsar, the pulsed energy spectrum varies with pulsar phase.  A study of the EGRET data by Fierro et al. (1998) of the phase-resolved emission of the three brightest gamma-ray pulsars (Vela, Geminga, Crab) showed no simple pattern of variation of the spectrum with phase that applied to all three pulsars. A broadband study of the Crab by Kuiper et al. (2001) indicated the presence of multiple emission components, including one that peaks in the 0.1 - 1 MeV range for the bridge emission between the two peaks in the light curve. Improved measurements and modeling of the phase-resolved spectra of pulsars can be expected to be a powerful tool for study of the emission processes. 

\end{itemize}

The measured spectra can be integrated to determine the energy flux of each pulsar. Except for the Crab and PSR B1509$-$58, whose luminosity peaks lie in the $\sim$100 KeV - 1 MeV range, the energy flux for the other gamma-ray pulsars is dominated by the emission above 10 MeV. Even then some interpolation or extrapolation may be needed for some pulsars that do not have spectra measured over a broad energy band; therefore these flux values have significant uncertainty. The energy flux can be converted to an estimated luminosity by using the measured distance to the pulsar and an assumed emission solid angle.  For simplicity, I assume an emission into one steradian.  This value is unlikely to be the same for all pulsars but provides a simple reference point for comparison.  A significant uncertainty in such calculations is introduced by the distance estimate.  Only for the nearby Geminga, B0656+14, and Vela pulsars is the distance well determined by parallax measurements (Brisken et al., 2002, 2003; Caraveo et al., 1996, 2001).  Analysis of other distances based on the radio Dispersion Measure is less certain and is dependent on a model of the Galactic electron density.  The distances given here reflect the new NE2001 electron density model of Cordes and Lazio (2002).  In particular, this new model finds that B1055$-$52 and J0218+4232 are substantially closer than previous models.  Table 1 summarizes these results for the ten pulsars. 

\begin{table}
\centering
\caption{\label{4} Summary Properties of the Highest-Confidence and Candidate Gamma-Ray Pulsars}
\bigskip
\begin{tabular}{lrrccccc}
\hline\hline
 Name  & P & $\tau$ & $\dot E$ & F$_E$ & d &  L$_{HE}$  & $\eta$ \\
     & (s) & (Ky) & (erg/s) &  (erg/cm$^{2}$s) & (kpc) & (erg/s)
& (E$>$1 eV)\\
\hline
Crab & 0.033 & 1.3 & 4.5 $\times$ 10$^{38}$ & 1.3 $\times$ 10$^{-8}$ & 2.0 & 5.0 $\times$ 10$^{35}$ &  0.001 \\
B1509$-$58 & 0.150 & 1.5 & 1.8 $\times$ 10$^{37}$ & 8.8 $\times$ 10$^{-10}$ & 4.4 & 1.6 $\times$ 10$^{35}$ &  0.009 \\
Vela & 0.089 & 11 & 7.0 $\times$ 10$^{36}$ & 9.9 $\times$ 10$^{-9}$ & 0.3 & 8.6 $\times$ 10$^{33}$ &  0.001 \\
B1706$-$44& 0.102 & 17 & 3.4 $\times$ 10$^{36}$ & 1.3 $\times$ 10$^{-9}$ & 2.3 & 6.6 $\times$ 10$^{34}$ &  0.019 \\
B1951+32 & 0.040 & 110 & 3.7 $\times$ 10$^{36}$ & 4.3 $\times$ 10$^{-10}$ & 2.5 & 2.5 $\times$ 10$^{34}$ &  0.007 \\
Geminga  & 0.237 & 340 & 3.3 $\times$ 10$^{34}$ & 3.9 $\times$ 10$^{-9}$ & 0.16 & 9.6 $\times$ 10$^{32}$ &  0.029 \\
B1055$-$52 & 0.197 & 530 & 3.0 $\times$ 10$^{34}$ & 2.9 $\times$ 10$^{-10}$ & 0.72 & 1.4 $\times$ 10$^{33}$ &  0.048 \\
\hline
B1046$-$58 & 0.124 & 20 & 2.0 $\times$ 10$^{36}$ & 3.7 $\times$ 10$^{-10}$ & 2.7 & 2.6 $\times$ 10$^{34}$ &  0.013 \\
B0656+14  & 0.385 & 100 & 4.0 $\times$ 10$^{34}$ & 1.6 $\times$ 10$^{-10}$ & 0.3 & 1.3 $\times$ 10$^{32}$ &  0.003 \\
J0218+4232 & 0.002 & 460,000 & 2.5 $\times$ 10$^{35}$ & 9.1 $\times$ 10$^{-11}$ & 2.7 & 6.4 $\times$ 10$^{33}$ &  0.026 \\
\hline
\end{tabular}
\end{table}

\section*{Other Candidate Isolated Neutron Stars}

Several sources seen in gamma rays have characteristics that strongly resemble those of the known gamma-ray pulsars, but without evidence of gamma-ray pulsation.  Examples are:   
                           
\begin{itemize}
\item 
 3EG J1835+5918/RX J1836.2+5925 is a likely Isolated Neutron Star (INS) similar to Geminga, based on multiwavelength observations (Mirabal et al. 2001, Reimer et al., 2001; Halpern and Mirabal 2001; Halpern et al. 2002) that show an X-ray source with both a thermal and nonthermal component, a gamma-ray source with no variability, a hard spectrum and high-energy cutoff, and no prominent optical or radio counterpart.  No pulsations have yet been found at any wavelength. 
 
\item PSR J2229+6114 is a young radio and X-ray pulsar with a high spindown rate, found in an EGRET source error box (Halpern et al., 2001).  Analysis of its energetics makes it a plausible counterpart for the gamma-ray source.  A search for pulsations in the gamma-ray data was inconclusive (Thompson et al., 2001). 

\item 3EG J2020+4017 (Brazier et al, 1996) and 3EG J0010+7309 (Brazier et al., 1998) are positionally associated with supernova remnants $\gamma$-Cygni and CTA1, respectively.  The gamma-ray sources may have X-ray counterparts, are non-variable, and have flat spectra similar to other gamma-ray pulsars. No pulsations have been seen from these sources, however. 

\item A number of other young radio pulsars have been found in EGRET error boxes (Camilo et al. 2001; D'Amico et al. 2001; Torres, Butt and Camilo 2001). The search for radio pulsars in EGRET error boxes is an ongoing effort, particularly using the large number of new pulsars being found with the Parkes survey (Kramer et al., 2003). 

\end{itemize}

Some, perhaps many, of these objects are likely to be gamma-ray pulsars.  In light of the relatively large gamma-ray error boxes and the absence of any gamma-ray pulsations, however, I take the conservative approach of including only the ten sources with at least some evidence of pulsed gamma rays in further discussions.

\section*{Gamma-Ray Pulsars Compared to the General Pulsar Population}

The sample of ten gamma-ray pulsars, using both the established and candidate sources,  can be compared to other pulsars in terms some of the derived 
physical parameters.   Figure 5 is a distribution of pulsars as a function of their period 
and period derivative, derived from the new ATNF Pulsar Catalogue (http://www.atnf.csiro.au/research/pulsar/psrcat/).  The gamma-ray pulsars are shown as squares.  Also shown 
are some of the derived physical parameters.  The gamma-ray pulsars tend to be 
concentrated (with the exception of the one millisecond candidate) in a region with 
high magnetic field (but not magnetar-strength) - shown by the dashed lines,  and 
relatively young ages - shown by the solid lines.  All ten gamma-ray pulsars share a 
third characteristic, shown by the dotted line, of having the open field line voltage 
high compared to most pulsars.  This is not surprising, since the particles are being 
accelerated electromagnetically.

\begin{figure}[b!] % fig 5
\centerline{\epsfig{file=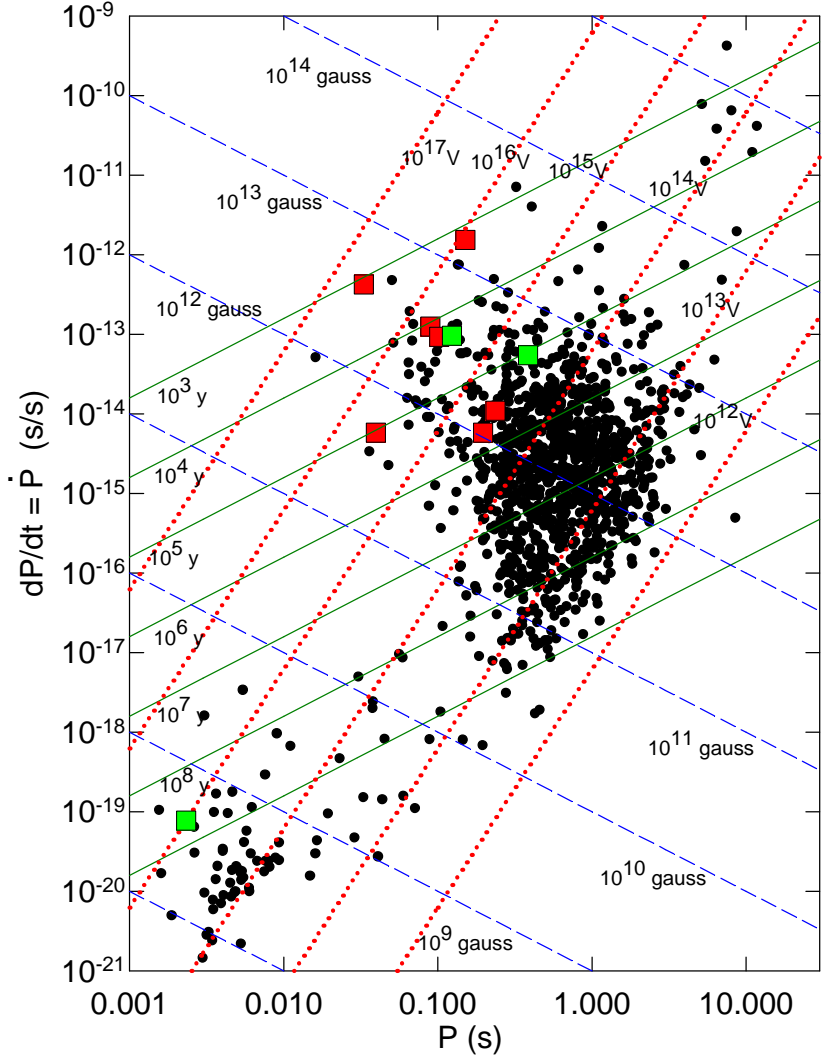,width=4.5in}}
\vspace{10pt}
\caption{Period v. period derivative for the ATNF Catalogue of Pulsars (http://www.atnf.csiro.au/research/pulsar/psrcat/). Small dots: no gamma-ray emission. Large dark boxes: seven high-confidence gamma-ray pulsars. Large light boxes: three lower-confidence gamma-ray pulsars. Solid lines: timing age. Dotted line: open field line voltage. Dashed line: surface magnetic fields.}
\label{fig5}
\end{figure}

\begin{figure}[b!] % fig 6
\centerline{\epsfig{file=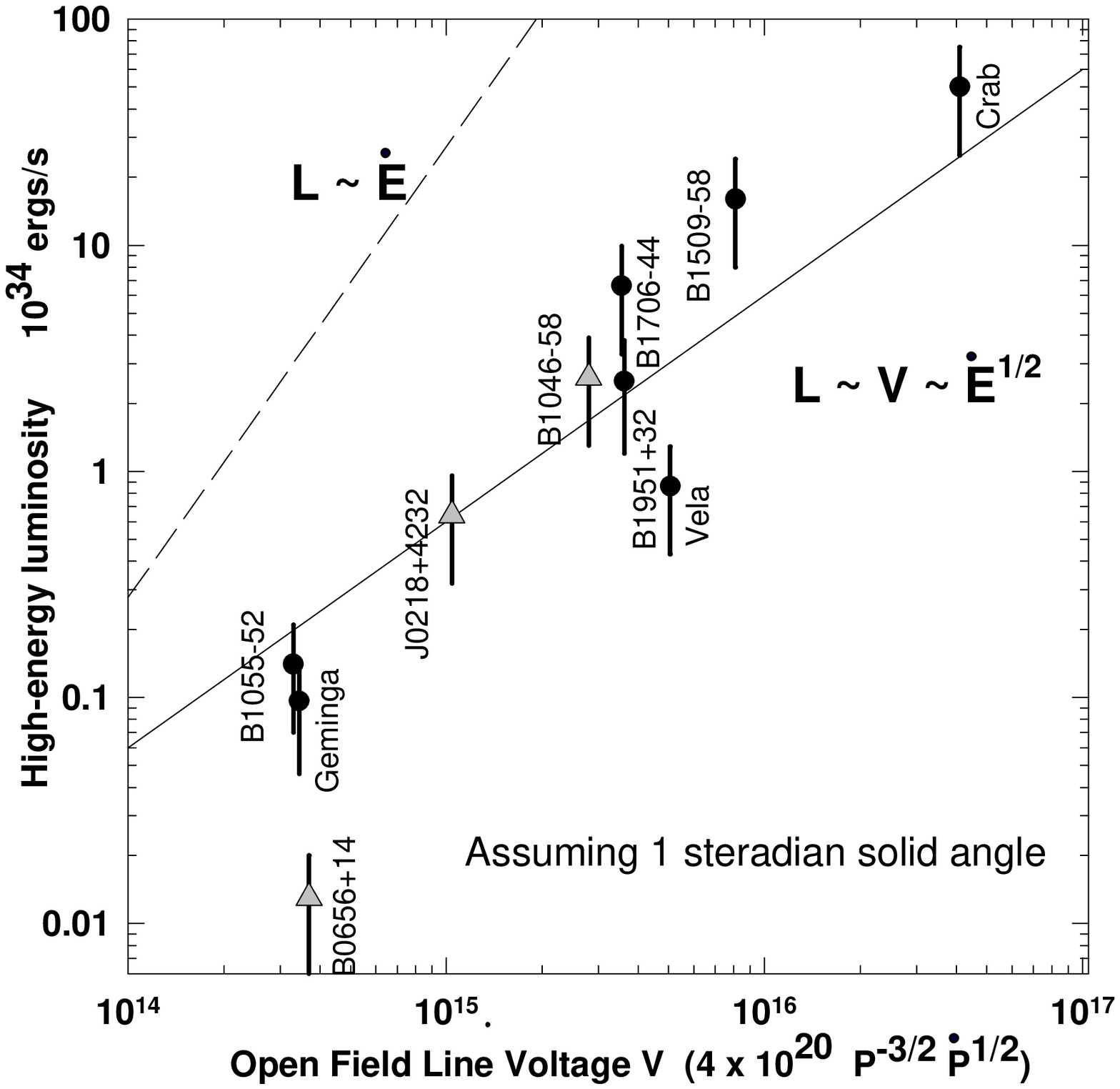,width=4in}}
\vspace{10pt}
\caption{High-energy pulsar luminosity as a function of the open field line voltage. Circles: high-confidence pulsars; Triangles: lower-confidence pulsars.}
\label{fig6}
\end{figure}

The open field line voltage, which is also proportional to the polar cap 
(Goldreich-Julian) current, has long been recognized as a significant parameter for 
gamma-ray pulsars (Arons, 1996).  Figure 6 illustrates that the pulsar high-energy luminosities, 
integrated above 1 eV, are approximately proportional to this parameter, shown by the 
solid line in the figure.  An interesting question is what happens for lower voltages, 
where the high-energy luminosity converges with the total spin-down energy 
available, shown by the dashed line.  This will be a question for future missions. 

\section*{Pulsars at the Highest Energies}

\begin{figure}[b!] % fig 7
\centerline{\epsfig{file=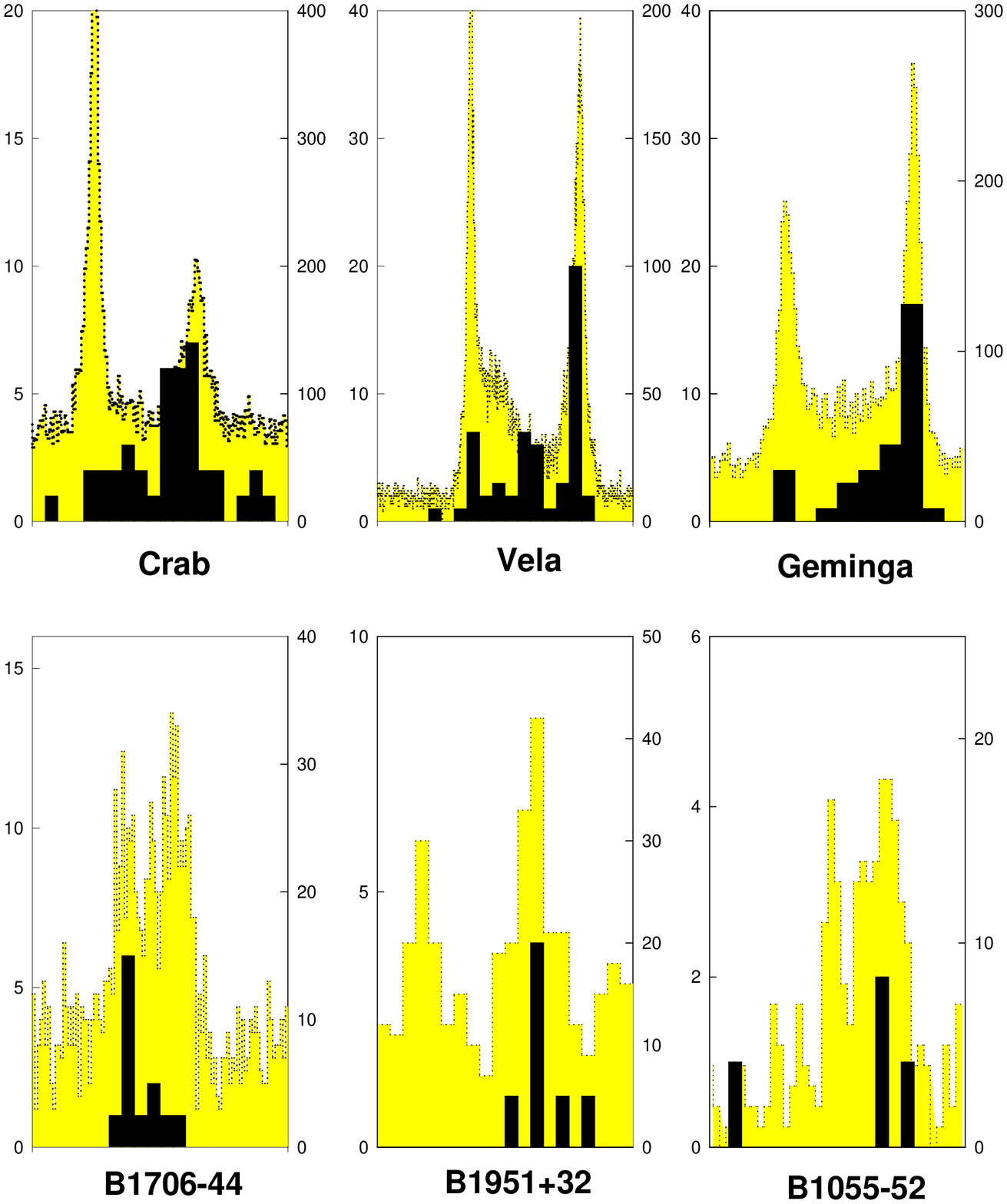,width=4.5in}}
\vspace{10pt}
\caption{Light curves of six gamma-ray pulsars above 100 MeV (dotted lines, right-hand scale) and above 5 GeV (dark histogram, left-hand scale). Each panel shows one full rotation of the neutron star.}
\label{fig7}
\end{figure}

	What happens to pulsars at the highest energies?  No pulsars have 
been seen at TeV energies. The upper end of the EGRET range represents the highest 
energies for detections of pulsed emission.  Figure 7 shows the 
light curves for six EGRET pulsars.   Despite the very limited statistics, several features of these pulsars are visible:
\begin{itemize}
\item 
 There is clear evidence for pulsed emission above 5 GeV for the Crab, Vela, Geminga, and B1706$-$44 pulsars, and the other two are consistent with having pulsed emission. 
 
\item In all cases, one of the two pulses seen at lower gamma-ray energies has faded at these high energies, leaving the other pulse dominant.  For all except PSR B1706$-$44, it is the trailing pulse that dominates (B1951+32 has pulses so close to 180$^\circ$ apart that determining the leading pulse is not clear). 

\item Except for the Crab, the emission from the other pulsars is concentrated strongly within the pulse as defined by the lower-energy gamma rays.  The extended emission from the Crab is thought to be unpulsed emission from the supernova remnant. 

\item The one pulse seen for the Crab appears to be displaced from the pulse seen at lower energies, a feature noted also by Kuiper et al. (2001), possibly a continuation of the narrowing of the gap between the two pulses that extends from the infrared to gamma-ray energies (Eikenberry and Fazio, 1997;  Ramanamurthy, 1994).  

\end{itemize}

\begin{figure}[b!] % fig 8
\centerline{\epsfig{file=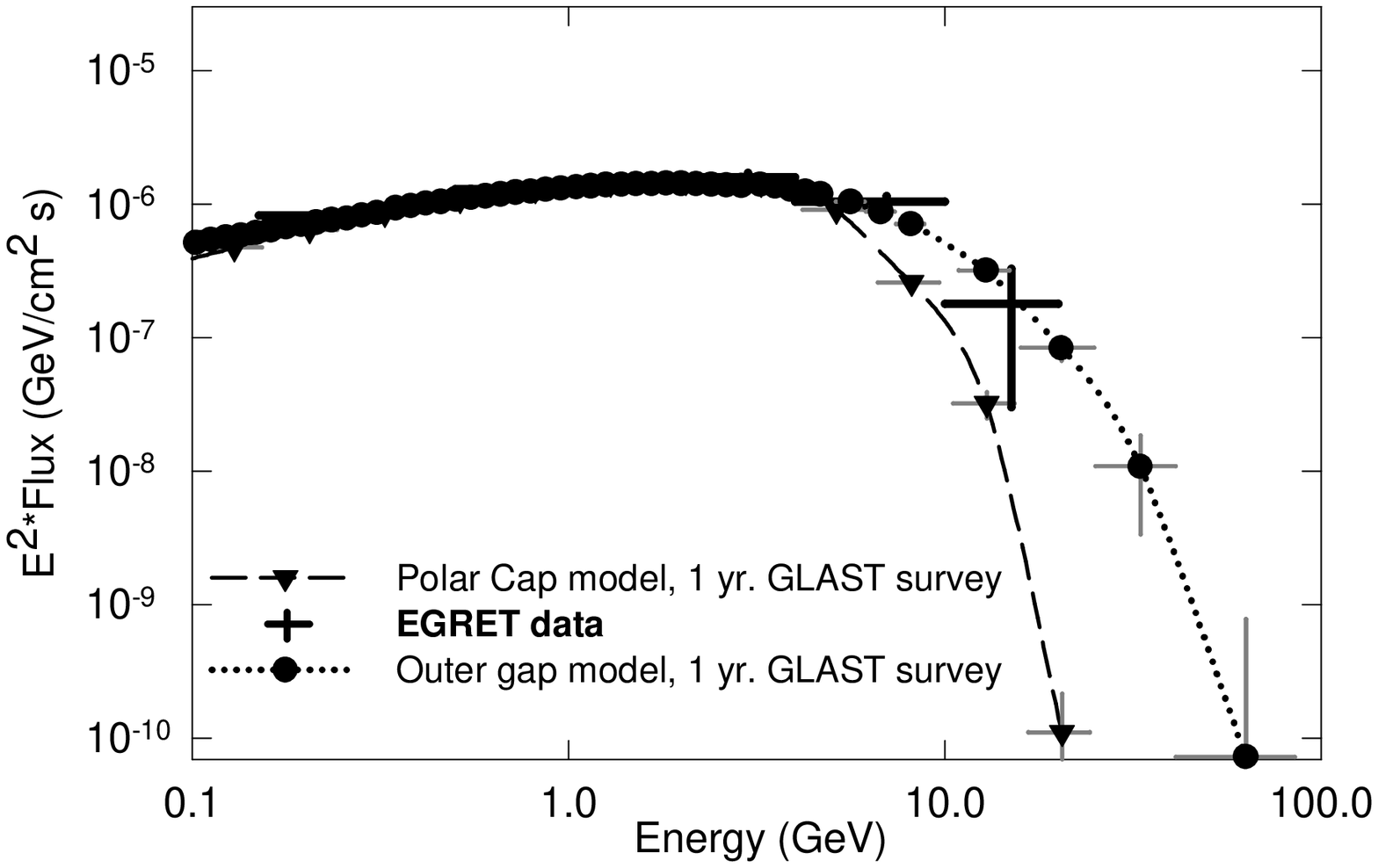,width=4in}}
\vspace{10pt}
\caption{High-energy spectrum of the Vela pulsar. Heavy error bars: EGRET data (Kanbach et al. 1994). Circles and dotted line: 
Outer gap model (Romani, 1996, 2000). Triangles and dashed line: Polar cap model (Daugherty and Harding 1996; Harding 2000). Error bars shown on the models are 
those expected from the GLAST mission (Gehrels and Michelson, 1999) in a one-year sky survey.}
\label{fig8}
\end{figure}

	The gap between about 10 GeV (where EGRET runs out of photons) and the 
current generation of ground-based telescopes is very important.  It is hard to predict 
from the limited EGRET data what is expected even at 100 GeV. Figure 6 shows the 
spectrum of Vela, the brightest of the EGRET pulsars. This spectrum is shown in $\nu$F$\nu$
or power per logarithmic energy interval format.  The 10-30 GeV point is based on 
only 4 photons, and the absolute calibration of EGRET at these energies is fairly 
uncertain. Also shown are the spectra expected in two popular gamma-ray pulsar 
models, the polar cap (Daugherty and Harding 1996; Harding 2000) and the outer gap (Romani, 1996, 2000).  The large error bars on the 
EGRET data make them consistent with both models.  The extrapolation to higher 
energies is, however, dramatically different.  There is also the possibility of a second, 
inverse-Compton component of the pulsed radiation at higher energies,  expected in 
some outer gap models (Romani, 1996), although there is yet no observational evidence for that  
component.  Searching for that second component will be an important goal of the 
next generation of very-high-energy gamma-ray telescopes.

\begin{figure}[b!] % fig 9
\centerline{\epsfig{file=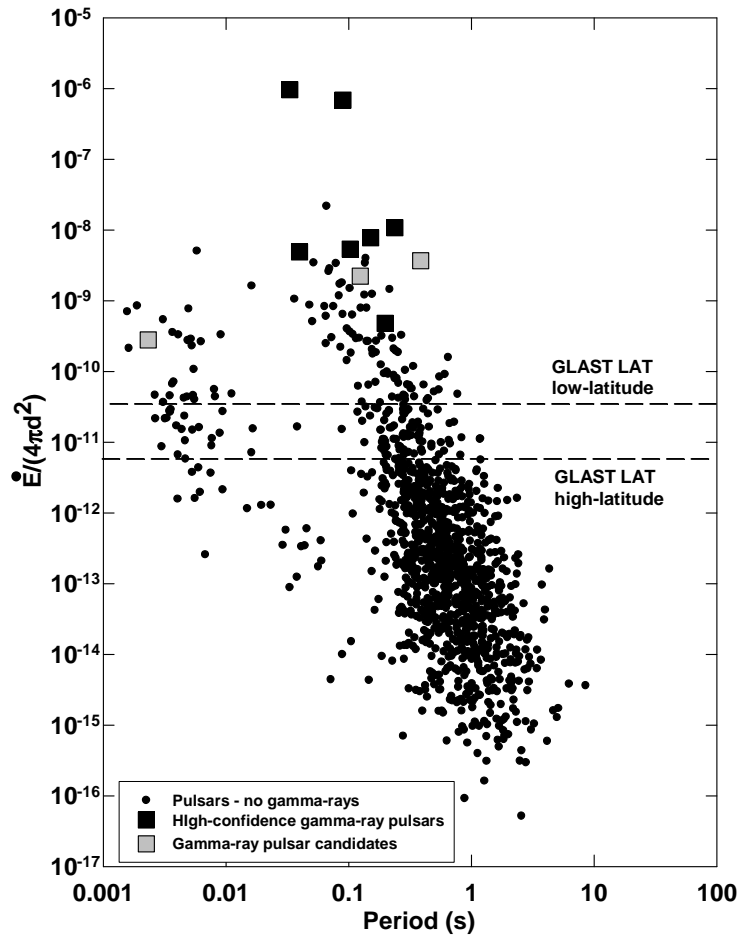,width=4in}}
\vspace{10pt}
\caption{Gamma-ray pulsar observability, as measured by the spin-down energy seen at earth.}
\label{fig9}
\end{figure}

\section*{Future Pulsar Observations at High Energies}

The high luminosity in the GeV energy range and the drop-off above that part 
of the spectrum suggest that the GLAST Large Area Telescope (LAT) will be the next big step in gamma-ray pulsar 
observations. The Italian AGILE mission will certainly make a contribution, especially in confirming 
those candidate pulsars for which the EGRET data are marginal, but GLAST LAT will be 
needed for a major increase in sensitivity and energy reach.  Figure 8 shows one of the 
ways that GLAST will help - the smaller error bars on the theoretical curves show two 
models' predictions folded through the GLAST LAT sensitivity for a one-year sky survey. 
GLAST LAT will easily distinguish spectra such as these and might detect a second 
component at higher energies.
A second strength of GLAST LAT will be in its much higher sensitivity than 
previous instruments.  Figure 9 shows one of the classic measures of pulsar 
observability: the pulsar spin-down luminosity divided by 4$\pi$ times the square of the 
distance, the total available pulsar energy output at Earth.  The 10 gamma-ray pulsars 
and candidates are shown as the large squares.  Six of the seven pulsars with the 
highest value of this parameter are gamma-ray pulsars. Below these, only a handful of 
pulsars are visible in gamma rays.  The GLAST LAT sensitivity will push the lower limit 
down substantially farther.  Two sensitivity limits are shown for GLAST LAT - one for 
low-Galactic-latitude sources and one for those at high latitudes, because the high 
diffuse emission along the Galactic plane reduces the sensitivity for point source 
detection.  The phase space that GLAST LAT opens up is substantial.

\begin{figure}[b!] % fig 10
\centerline{\epsfig{file=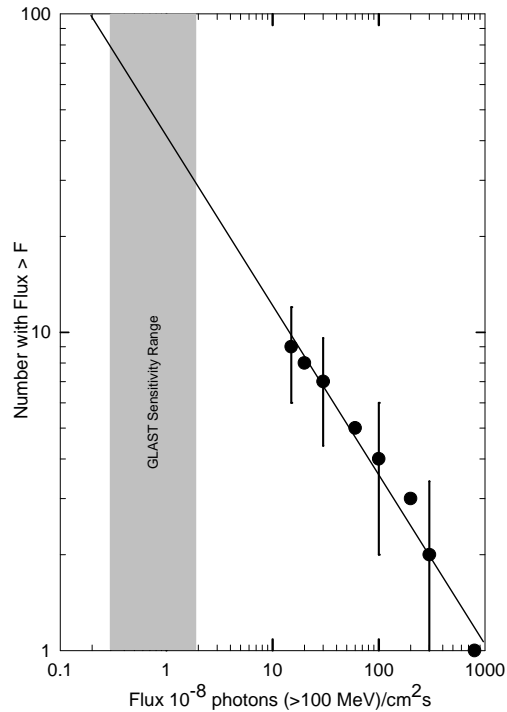,width=3in}}
\vspace{10pt}
\caption{Cumulative number of gamma-ray pulsars seen by EGRET as a function of 
observed flux.  Error bars are statistical and are shown on alternating data points for clarity.}
\label{fig10}
\end{figure}

How many pulsars will GLAST LAT see?  To some extent that number depends on 
which model best describes the emission and how the pulsars are distributed on the 
sky.  One empirical estimate can be made by constructing a logN-logS curve from the 
known pulsars.   Figure 10 uses all nine of the high-energy gamma-ray pulsars to 
construct this function.  The simple linear fit suggests that GLAST might expect to 
detect between 30 and 100 gamma-ray pulsars.  The range of sensitivity is dependent 
on where the pulsars lie with respect to the Galactic plane.  In some respects this 
figure is pessimistic, possibly due to the small sample.  A disk population should 
increase linearly with decreasing sensitivity,  but the fitted line goes as the $-$0.5 power 
of the sensitivity.  A linear function would raise the number of pulsars to more than 
100. Other methods have given even larger numbers of potential pulsar detections (e.g. McLaughlin and Cordes, 2000).

\begin{table}
\caption{Predicted gamma-ray fluxes (units of 10$^{-8}$ ph [E$>$ 100 MeV] cm$^{-2}$s$^{-1}$)}
\label{table1}
\begin{tabular}{lrrrr}
Pulsar J &  Zhang \& Harding & Rudak \&Dyks & Romani  &  Cheng \&Zhang    \\
   & (2000) &  (1998)  &   (2000)   &  (1998)  \\
\hline
 1932+1059  & $<$580$^a$ & 90 & $\sim$0 &  $<$16  \\
 0953+0755  & $<$500$^a$ & 90 & $\sim$0$^b$ &  $<$16  \\
 0437$-$4715  & $<$230$^a$ & $<$2 & $\sim$0 &  $<$16  \\
 2043+2740  & 50 & 30 & $\sim$0$^b$ &  50  \\
 1826$-$1334  & 20 & 20 & $\le$1 &  $<$16  \\
 1803$-$2137  & 20 & 20 & 30 &  $<$16  \\
 0742$-$2822  & 20 & 20 & $\le$7 &  25  \\
 0117+5914  & 20 & 20 & $\le$35 &  20  \\
 1801$-$2451  & 15 & 10 & 23 &  $<$16  \\
 1908+0734  & 50 & 10 & $\sim$0 &  $<$16  \\
 1730$-$3350  & 10 & 10 & $\le$20 &  $<$16  \\
 0538+2817  & 15 & 10 & $\le$24 &  20  \\
 0358+5413  & 10 & 8 & $\le$9 &  16  \\
 1453$-$6151  & 10 & $<$2 & $\sim$0$^b$ &  20  \\
 2337+6151  & 8 & $<$2 & $\le$11 &  20  \\
 1824$-$2452  & 16 & $<$2 & $\le$9 &  $<$16  \\
 1637$-$4553  & 4 & $<$2 & $\le$7 &  $<$16  \\
\hline
\end{tabular}
\begin{tabular} {l}
$^a$ Upper limit because the predicted emission is a large fraction of the pulsar spin-down luminosity.\\
$^a$ Small flux predicted due to beaming. 
\end{tabular}
\end{table}

Just as interesting as the number of pulsars to be seen by GLAST is which 
particular pulsars will be seen.  Such observations are probably the best discriminator 
among models. Table 2 is a compilation with help from the theorists involved, 
showing the expected flux for various radio pulsars that have not yet been seen in 
gamma rays. In some case, the predicted flux values exceed the limits seen with EGRET, indicating a limitation of the models. For many of these pulsars, different models make predictions that differ by 
more than an order of magnitude.  In these units, the GLAST LAT sensitivity is 2 for 
pulsars near the plane and 0.2 for high-latitude pulsars; therefore GLAST LAT will 
certainly provide solid observational tests of these and other gamma-ray pulsar 
models.

The third, and perhaps most important, capability of GLAST LAT for pulsars will 
be its ability to find unknown pulse periods, in order to look for more Geminga-like 
pulsars.  Various analyses have shown that only the brightest pulsars could be found in 
the EGRET data without independent information (Brazier and Kanbach 1996; Jones, 1998; Chandler et al. 2001). For all of the other 
unidentified EGRET sources, the photons are just too few and too far apart in time to derive 
unambiguous pulse periods.  With GLAST LAT, periodicity searches will be feasible for 
most, if not all, the unidentified EGRET sources (Mattox, 2000; McLaughlin and Cordes, 2000).  The potential is to find a whole 
new population of rotation-powered pulsars, much as X-ray astronomy has started to 
do in the past few years.  This could be an important new window onto the physics of 
the extreme conditions around these spinning neutron stars.

\section*{Summary}

\begin{itemize}
\item Gamma-ray pulsars are multiwavelength objects that provide a valuable probe of particle acceleration and 
interaction in the extreme conditions found near rotating neutron stars.

\item At least 7 pulsars are seen in gamma rays (six of those at energies above 100 MeV), with 3 or more additional good candidates and some other likely isolated neutron stars that have not yet shown evidence of pulsation.

\item The gamma-ray light curves generally differ from those at lower energies.  The double-pulse structure indicates emission above a single magnetic pole.  Above 5 GeV, the known gamma-ray pulsars show only one prominent peak in their light curves. 

\item The broadband energy spectra of these pulsars show multiple components.  The maximum power per frequency interval is found in the gamma-ray band, and all these pulsars show a high-energy cutoff, possibly related to the surface magnetic field strength. 

\item Compared to the general pulsar popultion, gamma-ray pulsars tend to be younger, have higher magnetic fields, and have higher open field line voltages.  The gamma-ray pulsar luminosities seem to increase with open field line voltage or particle current. 

\item The changing shape of the light curves and energy spectra in the 1-20 GeV range 
make this band particularly interesting for future observations.

\item GLAST, along with other satellite and ground-based gamma-ray telescopes, will 
make a major advance in gamma-ray pulsar studies.  Of particular interest for testing pulsar models will be learning which radio pulsars are gamma-ray bright, since different models make substantially different predictions. 

\end{itemize}


\begin{references} 
\bibitem{} Arons, J., {\it A\&AS} {\bf 120}, 49-60 (1996).
\bibitem{} Brazier, K.T.S. et al., {\it MNRAS} {\bf 281}, 1033-1037  (1996). 
\bibitem{} Brazier, K.T.S., Reimer, O., Kanbach, G., Carrima{\~n}ana, A., {\it MNRAS} {\bf 295}, 819-824 (1998). 
\bibitem{} Brazier, K.T.S. and Kanbach, G., {\it A\&AS} {\bf 120}, 85-87 (1996).
\bibitem{} Brisken, Walter F., Benson, John M., Goss, W. M., Thorsett, S. E., {\it ApJ} {\bf 571}, 906-917 (2002).
\bibitem{} Brisken, Walter F., Thorsett, S. E., Golden, A., Goss, W. M.,  {\it ApJ, submitted}, astro-ph/0306232 (2003).
\bibitem{} Caraveo, P. A., De Luca, A., Mignani, R. P., Bignami, G. F., {\it ApJ} {\bf 561}, 930-937 (2001).
\bibitem{} Caraveo, Patrizia A., Bignami, Giovanni F., Mignani, Roberto, Taff, Laurence G., {\it ApJ} {\bf 461}, L91-L94 (1996).
\bibitem{} Chandler, A. M. et al., {\it ApJ} {\bf 556}, 59-69 (2001).
\bibitem{} Cheng, K.S. and Zhang, L., {\it ApJ} {\bf 498}, 327-341 (1998).
\bibitem{} Daugherty, J. K. and Harding, A. K.,{\it ApJ} {\bf 458} , 278-292 (1996).
\bibitem{} Eikenberry, Stephen S., Fazio, Giovanni G., {\it ApJ} {\bf 476}, 281-290 (1997).
\bibitem{} Fierro, J. M., Michelson, P. F., Nolan, P. L., Thompson, D. J., {\it ApJ} {\bf 494}, 734-746 (1998).
\bibitem{} Finley, J.P., in: {\it The Soft X-Ray Cosmos}, ed. R. Petre, E.M. Schlegel, New York: AIP, AIP 
Conf. Proc. 313, 41-50 (1994).
\bibitem{} Gehrels, N. and Michelson, P., {\it Astroparticle Phys.} {\bf 11}, 277-282 (1999).
\bibitem{} Gotthelf, E. V., Halpern, J. P., Dodson, R., {\it ApJ} {\bf 567}, L125-L128 (2002).
\bibitem{} Halpern, J. P. et al., {\it ApJ} {\bf 552}, L125-L128   (2001).
\bibitem{} Halpern, J.P., Gotthelf, E.V., Mirabal, N., Camilo, F., {\it ApJ} {\bf 573}, L41-L44 (2002).
\bibitem{} Halpern, J.P. and Mirabal, N., {\it ApJ} {\bf 547}, L137-L140   (2001).
\bibitem{} Harding, A.K, private communication (2000).	
\bibitem{} Hartman, R. C. et al., {\it ApJS} {\bf 123}, 79-202   (1999).
\bibitem{} Jackson, M. S., Halpern, J. P., Gotthelf, E. V., Mattox, J. R., {\it ApJ} {\bf 578}, 935-942 (2002).
\bibitem{} Jones, B.B., PhD Thesis, Stanford University (1998).
\bibitem{} Kanbach, G. et al., {\it A\&A} {\bf 289}, 855-870 (1994).
\bibitem{} Kaspi, V.M. et al., {\it ApJ} {\bf 528}, 445-453 (2000). 
\bibitem{} Kramer, M. et al., {\it MNRAS} {\bf 342}, 1299-1324 (2003). 
\bibitem{} Kuiper, L. et al., {\it A\&A,} {\bf 351}, 119-132 (1999).
\bibitem{} Kuiper, L. et al., {\it A\&A} {\bf 359}, 615-626 (2000).
\bibitem{} Kuiper, L., et al., {\it A\&A} {\bf 378}, 918-935 (2001)
\bibitem{} Kuiper, L., Hermsen, W., Verbunt, F., Ord, S., Stairs, I., Lyne, A., {\it ApJ} {\bf 577}, 917-922 (2002).
\bibitem{} Kuz'min, A. D. et al., {\it A\&AS} {\bf 127}, 355-366 (1998).
\bibitem{} Kuz'min, A. D., Losovskii, B. Ya., {\it Astronomy Letters}, {\bf 23} 283-285 (1997).
\bibitem{} Mattox, J.R., private communication (2000).
\bibitem{} McLaughlin, M. A., Mattox, J. R., Cordes, J. M., Thompson, D. J., {\it ApJ} {\bf 473}, 763-772  (1996).
\bibitem{} McLaughlin, M. A., Cordes, J. M., {\it ApJ} {\bf 538}, 818-830  (2000).
\bibitem{} Mirabal, N., Halpern, J.P., Eracleous, M., Becker, R.H., {\it ApJ} {\bf 541}, 180-193  (2001).
\bibitem{} Nolan, P. L. et al., {\it ApJ} in press (2003).
\bibitem{} Ramanamurthy, P. V., {\it A\&A} {\bf 284}, L13-L15 (1994).
\bibitem{} Ramanamurthy, P.V. et al., {\it ApJ} {\bf 458}, 755-760 (1996).
\bibitem{} Reimer, O., et al., {\it MNRAS} {\bf 324}, 772-780 (2001). 
\bibitem{} Romani, R.W., {\it ApJ} {\bf 470}, 469-478 (1996).
\bibitem{} Romani, R.W., private communication (2000).
\bibitem{} Rudak, B., and Dyks, J., {\it MNRAS} {\bf 295}, 337-343 (1998). 
\bibitem{} Thompson, D. J., in {\it High Energy Gamma-Ray Astronomy}, American Institute of Physics (AIP) Proceedings, {\bf 558}, Edited by Felix A. Aharonian and Heinz J. V\"olk. American Institute of Physics, Melville, New York,  p.103-114 (2001).
\bibitem{} Thompson, D.J., {\it Adv. Space Res.} {\bf 25}, 659-668 (2000).
\bibitem{} Thompson, D.J. et al., {\it ApJ} {\bf 516}, 297-306 (1999).	
\bibitem{} Torres, D. F., Romero, G. E., Combi, J. A., Benaglia, P., Andernach, H., Punsly, B., {\it A\&A} {\bf 370}, 468-478  (2001).
\bibitem{} Torres, Diego F., Butt, Yousaf M., Camilo, Fernando, {\it ApJ} {\bf 560}, L155-L158 (2001).
\bibitem{} Torres, Diego F., Nuza, Sebastián E., {\it ApJ} {\bf 583}, L25-L29 (2003).
\bibitem{} Zhang, B. and Harding, A.K., {\it ApJ} {\bf 532}, 1150-1171 (2000).
\end{references}
\end{document}